\documentclass[aps,pre,11pt,a4paper]{revtex4-2}
\usepackage{amsmath,amsfonts,amssymb,amsthm,color}
\usepackage{graphicx}
\usepackage[english]{babel}
\usepackage{natbib}
\usepackage{physics}
\usepackage{enumitem}
\usepackage{mathtools}
\usepackage{cancel}
\usepackage[T1]{fontenc}

\parskip=\medskipamount

\newcommand{\eq}[1]{(\ref{#1})}
\newcommand{\fig}[1]{Fig.~\ref{#1}}

\newcommand{\be}{\begin{equation}}
\newcommand{\ee}{\end{equation}}
\newcommand{\beq}{\begin{equation}}
\newcommand{\eeq}{\end{equation}}
\newcommand\disp{\displaystyle}

\makeatletter
\renewcommand{\fnum@figure}{Fig. \thefigure}
\makeatother

\newcommand{\mbf}{\mathbf}

\newcommand{\mcal}{\mathcal}

\newcommand{\lamb}{\lambda}

\begin{document}

\title{Motif enrichment as a driver of scale-free behavior in rewired random
regular graphs}

\author{Pawat Akara-pipattana$^{1}$, Sergei Nechaev$^{1,2}$, and Bhargavi Srinivasan$^{1}$}
\affiliation{$^1$LPTMS, CNRS -- Universit\'e Paris Saclay 91405 Orsay Cedex, France \\ $^2$Beijing Institute of Mathematical Sciences and Applications (BIMSA), Yanqi Lake, Huairou District, Beijing 101408, China}

\begin{abstract}

We study the statistics of rewired random regular graphs (RRGs) in a mixed ensemble, where the average number of triangles is controlled by the fugacity $\lambda$, while the number of vertices and the vertex degree are fixed. This model exhibits a phase transition at critical fugacity $\lambda_{cr}$ from a triangle-poor phase (TPP) to a triangle-rich phase (TRP). We estimate $\lambda_{cr}$ by comparing the entropy of TPP with the energy of TRP. Above $\lambda_{cr}$, the RRG becomes a two-phase system in which dense clusters are connected by a sparse inter-cluster subgraph. Numerically the clusters and the inter-cluster subgraph are identified through the Ollivier-Ricci curvature of the edges: it is positive inside clusters and negative on the links that form the inter-cluster subgraph. The inter-cluster subgraph is characterized by a degree distribution, $P(d) \sim d^{-\gamma}$, with $\gamma \approx 1.5$. We attribute this behavior to an "emergent preferential attachment" induced by triangle motifs. A kinetic growth model suggests that the attachment leads to the rate of growth superlinear in vertex degree. We find that most inter-cluster triangles are isosceles, with the base lying inside one cluster and the apex belonging to the inter-cluster subgraph. It is conjectured that a spanning inter-cluster scale-free subgraph is stabilized against complete fragmentation into a disconnected set of cliques, by a sparse, structurally rigid "gas" of such apex-shared isosceles triads.

\end{abstract}

\maketitle

\section{Introduction}
\label{sect:1}

Networks with nontrivial topological organization are ubiquitous in complex systems. They emerge in physics, biology, sociology, and many other domains of science, technology, and everyday life. One of the central questions in the study of such systems concerns the appearance of scale-free structures, characterized by a vertex degree distribution with a power-law tail, $P(d) \sim d^{-\gamma}$, where $\gamma$ typically lies in the range $\gamma\in [2,3]$ -- see, for example, \cite{PastorSatorras2001, Newman2005, Lee2005, Song2005}. Such a power-law distribution signals the presence of hubs and hierarchical organization, qualitatively distinct from that of classical random graph models such as Erd\H{o}s–R\'{e}nyi graphs (ERGs), or random regular graphs (RRGs). 

Another important feature distinguishing artificial and functionally oriented “physical” networks from abstract random graphs, is the existence of local motifs. From a mathematical perspective, motifs such as triangles are the simplest closed configurations. Their presence signals clustering, commonly quantified by the clustering coefficient, and is associated with a functional robustness in biological reaction networks, or correlations between agents sharing a common neighbor in social networks. In ERGs, the expected number of spontaneously formed triangles scales as $\binom{N}{3}p^3$, where $p$ is the connection probability of two arbitrary vertices (nodes) of the graph, and $N$ is the number of nodes. In the sparse-graph limit, where $p\sim d/N$ with $d$ being the average vertex degree, this quantity is of order one, leading to a vanishing fraction of triangles per vertex in the thermodynamic limit $N\to\infty$. In contrast, real-world networks often exhibit clustering that far exceeds expectations for random graphs -- see, for example, \cite{Chen2017,PMC3151270}. 

The first attempt to design random graphs with controlled triangular motifs for modeling clustering phenomena, can be traced back to the work of Strauss \cite{strauss}, which has been further generalized in \cite{tunneling, Kenyon_2017-1, Kenyon_2017-2,Gorsky:2020cnaa008, newman, burda}. In these works triangles in Erd\H{o}s–R\'{e}nyi graphs are introduced via a fugacity term. However, this approach fails to produce clustering in the low connectivity regime. An attempt to resolve this issue was proposed by Grassberger \cite{grassberger} in the "biased rewiring model", where the "soft" constraints on vertex degrees in the Strauss model was replaced by the "hard" ones, while keeping the control over the average number of triangles with a fugacity term. This permitted control independently the clustering and the degree distribution. Such an ensemble can be thought of as a graph satisfying the degree constraint, which is then rewired to increase the number of triangles. The rewiring process is applied until the number of triangles reaches a steady state corresponding to the fugacity. The degree constraints preserve the number of edges during the rewiring process. Some probabilistic and computational aspects of rewired ensembles have been discussed in \cite{maslov, viger}.

The rewiring of ERGs with a conserved vertex degree distribution was studied in a series of works \cite{tunneling, pandemic, new-localization, massaction}, where it was shown that the prevalence of triangles correlates with changes in global statistical properties of the ensemble. Above certain value of fugacity of triangles, the rewiring results in clustering and developing an inter-cluster subgraph with a power-law degree distribution \cite{pandemic}. The transition is accompanied by the change in the spectral density of the adjacency matrix which acquires a triangular shape \cite{tunneling}. This links two graph properties together: geometry (manifested in clustering), and new emergent scale invariance (manifested in the scale-free behavior of the inter-cluster subgraph). 

\begin{figure}[ht]
    \centering
    \includegraphics[width=\linewidth]{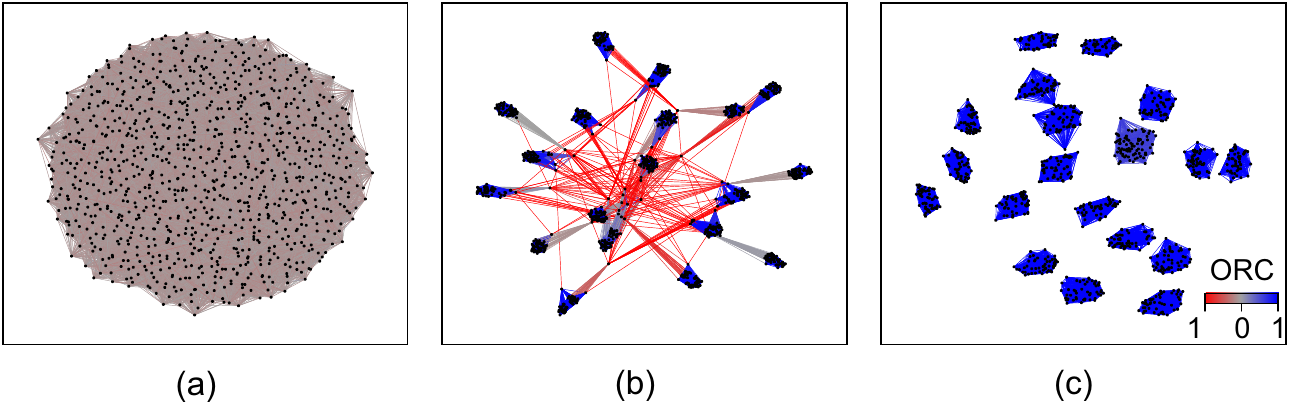}
    \caption{Different configurations of RRGs with $N=1000$ and $d=50$, whose edges are colored according to their Ollivier-Ricci curvature (ORC) sorted by the number of triangles: (a) typical RRG in a triangle-poor phase with negative ORC; (b) loosely connected clusters in a triangle-rich phase consisting of clusters with a positive ORC (blue), connected by the inter-cluster subgraph with a negative ORC (red); (c) disconnected cliques configuration with the maximum number of triangles and positive ORC (blue).}
    \label{fig:intro}
\end{figure}

In this work, we further explore this connection by studying the rewiring process in random regular graphs (RRGs). Recall that a $d$-regular random graph is chosen uniformly at random from the ensemble of all graphs on $N$ vertices in which every vertex has a fixed degree $d$. Rewiring RRG has an advantage over ERG, as the "ground state" of RRG is known due to its homogeneous degree, $d$: the phase with the maximal number of triangles is represented by the collection of independent full subgraphs. Under degree-preserving rewiring, the RRG exhibits a transition at a critical triangle fugacity, from a homogeneous, triangle-poor phase (\fig{fig:intro}a) to an inhomogeneous, triangle-rich long-living state of loosely connected clusters (\fig{fig:intro}b). 

Clustered structure of the RRG in the triangle-rich phase does not have the highest number of triangles. The true ground state with the maximal number of triangles corresponding to the complete defragmentation of the graph into disconnected collection of cliques shown in \fig{fig:intro}c, is inaccessible for times of observation. A detailed discussion of the transition is provided in Section \ref{sect:2}. In \fig{fig:intro}b clusters are identified using the Ollivier-Ricci curvature (ORC) according to the algorithm explained in Section \ref{sect:2}, with all necessary definitions given in Appendix \ref{sect:app1}. 

Similarly to its ERG counterpart, RRG also demonstrates changes in spectral properties under rewiring, accompanied by the emergence of power-law degree distribution of the inter-cluster subgraph. This has previously been explored in \cite{Kochergin:2023prb} in the context of Anderson localization and multifractality.

We focus on the following questions: (i) how the average number of triangles depends on the fugacity in the triangle-poor phase; (ii) how to estimate the transition point between the triangle-poor and triangle-rich phases, and (iii) what is the origin of the scale-free behavior of the inter-cluster subgraph in the triangle-rich phase.

The paper is structured as follows. In Section \ref{sect:2}, we introduce the Gibbs ensemble of random graphs with a rigid constraint on vertex degree, thus defining an ensemble of RRGs, specify the rewiring process, and formalize the definition of the inter-cluster subgraph and clusters using Ollivier–Ricci curvature. In Section \ref{sect:3}, we explore rewired RRGs at different values of the fugacity, $\lambda$, of triangles. Below a critical $\lambda$, the number of triangles follows a law of mass action, corresponding to "chemical reactions" between triads of various types. Using entropic arguments, we estimate the transition from the triangle-poor to the triangle-rich phase. In Section \ref{sect:4}, we discuss the structure of inter-cluster subgraphs generated in the triangle-rich phase. We show that the inter-cluster vertex degree distribution follows a power law $P(d) \sim d^{-\gamma}$ with a typical scaling exponent $\gamma \approx 1.5$. We formulate the rewiring process as an effective (mean-field) growth dynamics. This approach reveals an emergent preferential attachment mechanism responsible for the scale-free nature of the inter-cluster subgraph with the growth rate having superlinear dependency on vertex degree. In Section \ref{sect:5}, we discuss the possible implications of our findings.

\section{Key definitions and the model}
\label{sect:2}

Consider a \emph{mixed} ensemble of random regular graphs, ${\cal G}$, with vertex degree $d$. The notion "mixed" means that the ensemble is microcanonical in number of vertices, $N$, and edges, while it is canonical in number of triangular motifs. The average number of triangles is controlled by the chemical potential $\lambda$, which has a sense of the fugacity (energy) of the triangle.

The partition function $Z$ of the ensemble of random regular $N$-vertex graphs, ${\cal G}$, characterized by the adjacency matrix $A$, can be formally written as
\be
Z = \sum_{\{A\}} e^{\frac{\lambda}{6}\tr A^3} \prod_{k=0}^N \delta\left(\sum_{l=0}^{N} A_{kl} - d\right),
\label{eq:01}
\ee
where $\sum_{\{A\}}$ means the summation over all adjacency matrices $A$ with matrix elements $A_{kl}$ and the vertex degree $d$, and $\lambda$ is the fugacity conjugate to the number of triangles $\tfrac{1}{6}\tr A^3$. We define $H=\tfrac{\lamb}{6}\text{tr}\mbf{A}^3$ as the energy of our model. This model, being similar to the Strauss-like model studied extensively in \cite{strauss, burda, newman}, has one important distinction: additional vertex degree constraints prevent the graph from collapsing into a single cluster, as observed in the dense phase of the Strauss model.

\begin{figure}[ht]
    \centering
    \includegraphics[width=0.9\linewidth]{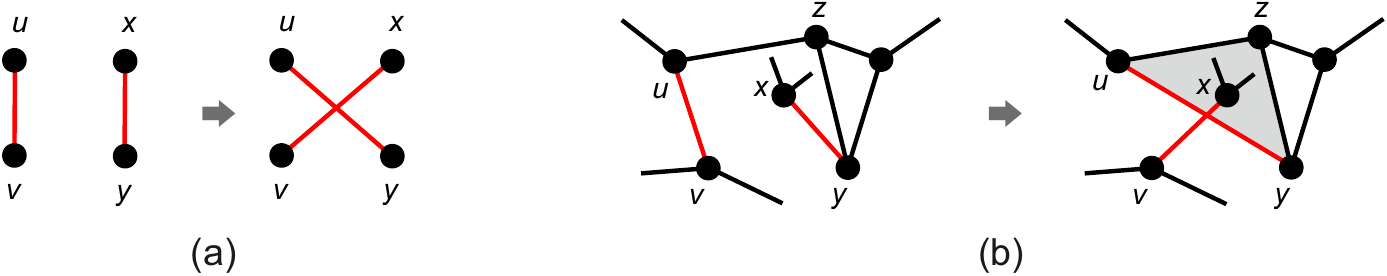}
    \caption{Maslov-Sneppen vertex degree-preserving rewiring: (a) Exchange of edges preserving vertex degree; (b) Sample of a graph in which the number of triangles in increased after one step of rewiring.}
    \label{fig:rewirings}
\end{figure}

Graphs in the ensemble are sampled using the Maslov-Sneppen degree-preserving rewiring process \cite{maslov} implemented through the Metropolis Monte Carlo algorithm described by repeating the sequence of the following steps:
\begin{enumerate}
\item We prepare an initial graph to be rewired. This initial state is either a generic RRG or the fully connected cliques. Different initial states yields into different steady state configurations under this rewiring process, as we will discuss in Section \ref{sect:3}.

\item At each Monte Carlo step, a rewiring move to replace an edge pair $(u,v)$ and $(x,y)$ by a pair $(u,y)$ and $(x,v)$ is proposed as it is shown in \fig{fig:rewirings}a. If any of the edges $(u,y)$ or $(x,v)$ is already occupied, the move is invalid and a new rewiring move to $(u,x)$ and $(v,y)$ is proposed instead. If any of the edges in this newly proposed move are also occupied, a new set of two vertex pairs is considered. This step is repeated until a valid rewiring move is found.

\item The proposed move is accepted with probability $\min(1,e^{\Delta H})$ where $\Delta H$ is the energy difference between the current and the proposed state, as prescribed by the conventional Metropolis algorithm. The rewiring move is always accepted if it increases the number of triangles in the graph as it is shown in \fig{fig:rewirings}b. Steps 2 and 3 of the algorithm together constitute a single Monte Carlo step of the degree-preserving rewiring process.

\item The rewiring process continues until the number of triangles reaches a steady state, while $\lamb$ is kept \emph{fixed} throughout the process.
\end{enumerate}

Since there is no insertion or removal of a new edge, all vertex degrees are conserved. For the parameters explored in this work, we find heuristically that it takes of order $N\times 10^6$ Monte Carlo steps  to reach the steady state.

The determination of a cluster is a delicate issue. Algorithms with external parameters, such as  modularity-based approaches \cite{newmanbook} can lead to parameter-dependent results. To make a clear distinction between the edges inside and outside of clusters, we use the concept of Ollivier-Ricci curvature (ORC) \cite{ollivier1}. ORC is a generalization of the notion of the curvature in graphs using the optimum transport concept through the Wasserstein distance \cite{Lin2011}. For completeness, we recall in Appendix \ref{sect:app1} the corresponding construction. The community detection using ORC has been considered in \cite{comm_detect,sia2019ollivier}. ORC has also been used to model geometrical structures in random graph models \cite{Trugenberger_2017,Trugenberger_2025}.

Conceptually, ORC compares the neighborhoods of the two endpoints of an edge: it measures the cost of transporting a unit of mass spread over the neighbors of one endpoint onto the neighbors of the other. It is negative on a bottleneck edge, whose two neighborhoods are linked only through the edge itself, so that the mass must travel across it, and positive when the neighborhoods overlap or are joined by short paths. For example, no edge of a tree has positive ORC, since any two neighborhoods meet only through a single edge. On the other hand, ORC is positive on the edges of a fully connected graph, where all neighbors are shared, and for a given degree it is largest there \cite{Lin2011,Srinivasan2025}. 

In our work, subgraphs with positive ORC edges and their vertices define clusters, while the connected set of edges and their vertices with negative ORC defines the inter-cluster subgraph. Some vertices are shared between clusters and inter-cluster subgraphs. However, each edge strictly belongs to either one of the clusters or the inter-cluster subgraph. Later in Section \ref{sect:4} when discussing the degree distribution of the inter-cluster subgraph, we take into account edges with the negative ORC only (see Appendix \ref{sect:app1} for details). Such a definition of a cluster is consistent with a more traditional cluster determination based on the clustering coefficient \cite{WattsStrogatz1998}: the more triangles share a common edge, the greater the positive value of ORC the edge acquires. We compute ORC using the Python package used in \cite{comm_detect}.

\section{Phases and critical behavior of triangle-enriched random regular graphs}
\label{sect:3}

Here we classify the triangle-rich and the triangle-poor phases. We provide an estimation for a number of triangles in the triangle-poor phase based on a mean-field model of "chemical reactions" among triads of different types \cite{massaction}. In addition, we show that the critical behavior of the rewired RRG ensemble is sensitive to the number of triangles in the initial state. In particular, configurations with initially large number of triangles undergo the transition between the triangle-poor and triangle-rich phases at a lower value of $\lamb$ than configurations with initially small number of triangles.

To begin with, consider two limiting cases: $\lamb=0$ and $\lamb\rightarrow \infty$. For $\lamb=0$, there is no bias towards the creation of triangles, and we have a standard RRG ensemble. The probability of a spontaneous triangle formation, given that triangles form independently, is $(\tfrac{d}{N})^3$. Thus, for $N\gg 1$, the expected number of triangles at $\lamb=0$ is 
\be
\triangle_{RRG} \equiv \lim_{N\rightarrow\infty}\binom{N}{3}\left(\frac{d}{N}\right)^3= \frac{d^3}{6}
\label{eq:tr_poor}
\ee
For $\lamb \to \infty$, one expects that the rewiring leads to configurations with maximal possible number of triangles. For RRGs of degree $d$, this leads to the disintegration of a graph into $\tfrac{N}{d+1}$ disconnected subgraphs, each being a fully connected cluster of $d+1$ vertices. This results in the following maximal number of triangles:
\be
\triangle_{max} = \frac{d(d-1)N}{6}.
\label{eq:max-triang}
\ee
We refer to this configuration as \emph{disconnected cliques}. For $N\gg 1$ the value $\triangle_{max}$, being $N$--dependent, is much larger than the number of triangles $\triangle_{RRG}$ in \eq{eq:tr_poor}. Thus, we can distinguish two phases of the graph ensemble depending on the scaling of the number of triangles with $N$: a triangle-poor phase (TPP), where the number of triangles is $N$-independent, and a triangle-rich phase (TRP), where it grows with $N$. We normalize the number of triangles by $\triangle_{\max}$, obtained from the rewiring procedure. 

\begin{figure}[ht]
\centering
\includegraphics[width=0.9\textwidth]{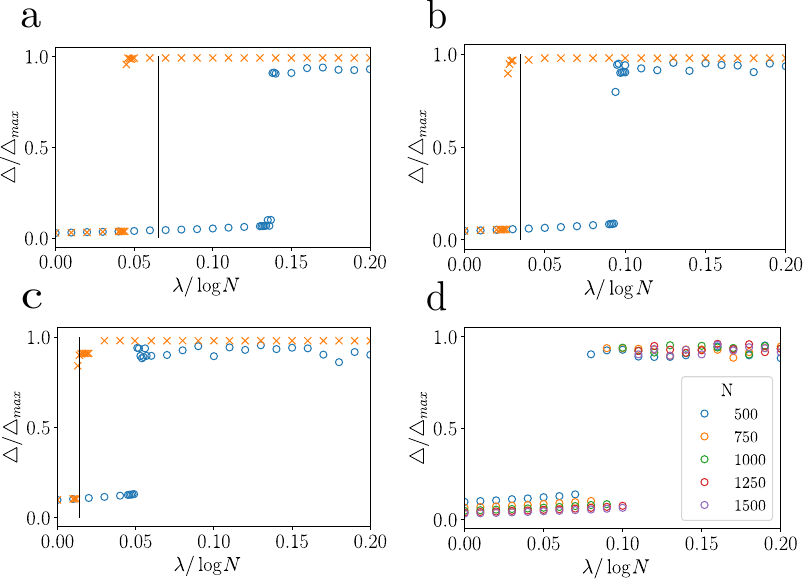}
\caption{Normalized number of triangles, $\triangle/\triangle_{\max}$, as a function of $\lamb/\log N$. Blue dots designate "forward rewiring" when we start with triangle-poor phase and increase $\lamb$. Orange dots correspond to "backward rewiring" when we start from disconnected cliques and decrease $\lamb$. Black vertical lines designate the estimated positions of transition points $\lamb_{cr}^-$, on the basis of \eq{eq:lamb_BC}. Panels (a)-(c) show transitions in RRGs with $N=1000$ and $d$ = 30, 50, 100 respectively. Panel (d) demonstrates how curves for the forward rewiring of RRG with $d=50$ collapse in coordinates with the rescaled fugacity $\lambda/\log N$ for various $N$.}
\label{fig:hyst}
\end{figure}

In \fig{fig:hyst}a-d we show how the normalized number of triangles, $\triangle/\triangle_{\max}$, depends on the normalized fugacity, $\lambda/\log N$. This choice of normalization for the fugacity follows from the scaling of transition points on $N$, as discussed later in Section \ref{sect:3.2}. 

We consider rewiring that begins at two different initial states. 
\begin{itemize}
\item[(i)] Starting at a typical RRG, the rewiring \emph{increases} the number of triangles; we refer this process as a "forward rewiring" from triangle poor phase (TPP) to triangle rich phase (TRP). The TPP $\to$ TRP trnsition occurs at a critical value $\lamb^+$; 
\item[(ii)] Starting at disconnected cliques corresponding to maximal TRP, the rewiring process \emph{decreases} the number of triangles. We call TRP $\to$ TPP rewiring as a "backward rewiring". The TRP $\to$ TPP transition occurs at a critical value $\lamb^-$, where $\lamb<\lamb^+$.
\end{itemize}

The steady state of a forward rewiring (blue circles) demonstrates two distinct behaviors:
\begin{enumerate}
\item In TPP ($\lamb$ is small), the number of triangles slowly increases with the fugacity, fully consistent with the mean-field estimates based on the law of mass actions for elementary chemical reactions discussed in Section \ref{sect:3.1}. Graphs in this phase have no apparent clusters and are invisible. The Ollivier-Ricci curvature is homogeneous in this phase as shown in \fig{fig:intro}a.

\item In TRP, ($\lamb$ is large), the number of triangles fluctuates near the maximal value independent of $\lamb$. Graphs in this phase are loosely connected clusters with a positive in-cluster ORC, while the inter-cluster subgraph has a negative ORC, as shown in \fig{fig:intro}b. The number of clusters in the TRP is approximately $N/d$. 

\end{enumerate}

\begin{figure}[ht]
    \centering
    \includegraphics[width=0.45\linewidth]{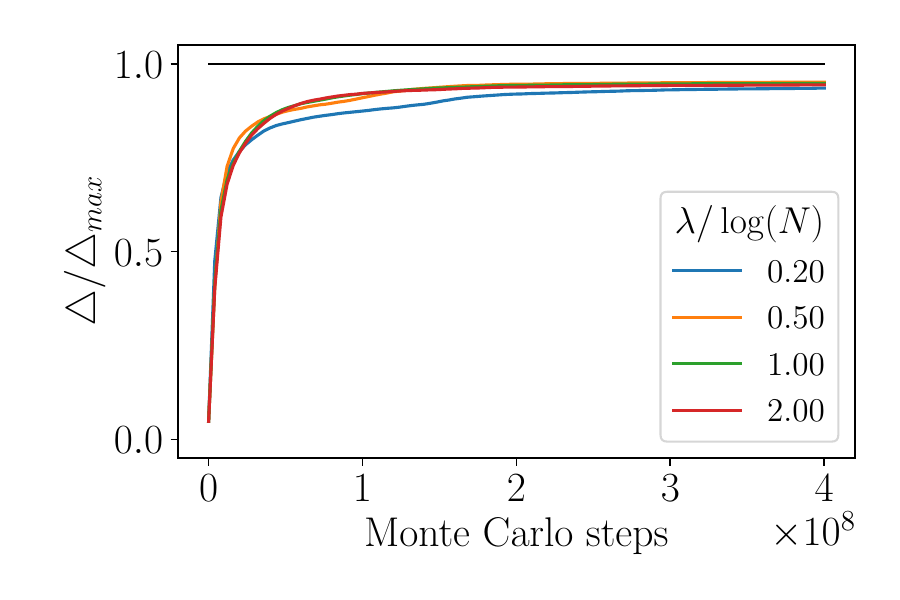}
    \includegraphics[width=0.45\linewidth]{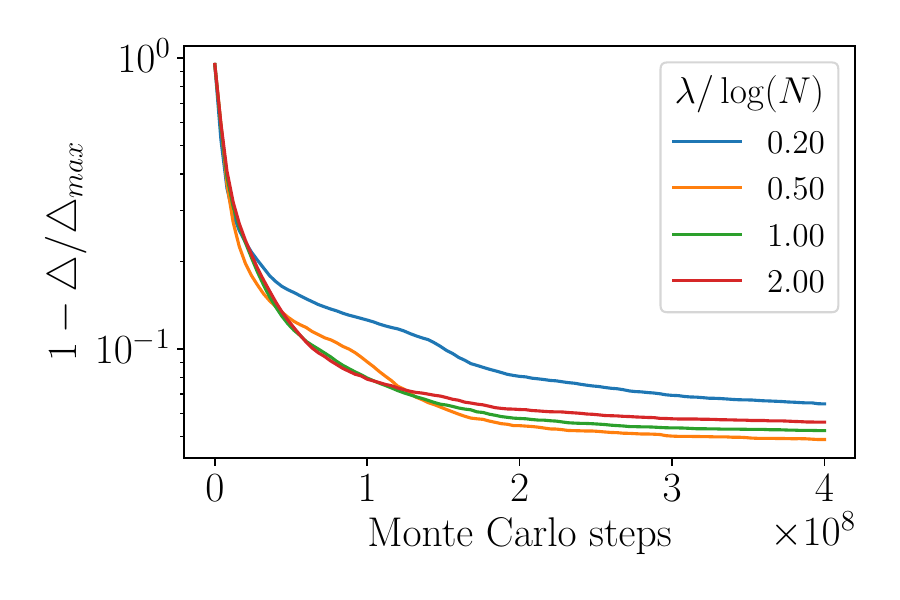}
    \caption{Left: the normalized number of triangles as a function of a number of Monte Carlo steps for rewiring of RRG with $N=1000$ and $d=50$ at $\lamb=0.2,0.5,1,2$. The gap between curves and the black horizontal line emphasizes that the maximum number of triangles is not reached. Right: the dependence of $1-\triangle/\triangle_{max}$ on number of Monte-Carlo steps in semi-log scale for better differentiation between the curves. Note that higher value of $\lamb$ do not necessary lead to the steady state with bigger number of triangles.}
    \label{fig:stab}
\end{figure}

Figure \ref{fig:hyst}d demonstrates that the normalized number of triangles $\triangle/ \triangle_{max}$ decreases with growing system size $N$, in TPP, while it remains of the same order of magnitude regardless the value of $N$ in TRP, in agreement with our classification of phases. In forward rewiring, the normalized number of triangles as a function of Monte Carlo steps above $\lamb^+$ reaches the quasi-stationary state corresponding to a configuration of loosely connected clusters. In \fig{fig:stab}, we show that clusters are kinetically trapped in a long-living state \emph{below} the complete fragmentation of the graph into the set of independent cliques. Increasing fugacity $\lamb$ allows the system to reach the steady state faster, but does not increase the number of triangles in the steady state. Fluctuations in the number of triangles suggest that these configurations are stable. In Section \ref{sect:5} we speculate about possible mechanism behind such stability. Similar "trapping" is also observed in other random graph models with hard constraints \cite{birth_geo}.

The transition between the triangle-rich and the triangle-poor phase is sensitive to the number of triangles at the preparation condition. In the backward rewiring, where we start from a set of disconnected cliques, designated by orange crossed in \fig{fig:hyst}a-c, we observe that the transition TRP $\to$ TPP occurs at a lower value of the fugacity, $\lambda^- < \lambda^+$. Moreover, for any $\lambda$ approaching $\lambda^-$ from above ($\lambda \searrow \lambda^-$), the number of triangles stays close to the maximum $\Delta_{max}$, and the graph consists of a collection of completely disconnected clusters. This contrasts with the forward rewiring TPP $\to$ TRP for $\lambda > \lambda^+$, where the graph becomes trapped in a long-living quasi-stationary state of loosely connected clusters.

\begin{figure}
    \centering
    \includegraphics[width=0.7\linewidth]{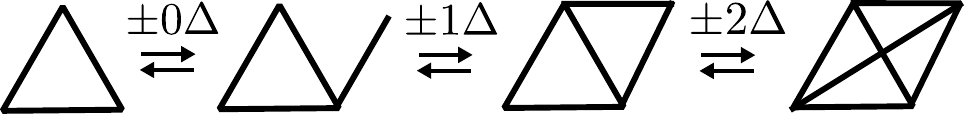}
    \caption{The creation of a clique of 4 vertices from a clique of 3 vertices happens in 3 steps. In the first step, an added edge does not increase any number of triangles, making it relatively unfavorable. However, the advantage of forming triangle increases sequentially at each step. On the other hand, the first step to destroy a clique of 4 vertices is also unfavorable since it reduces the number of triangle by 2. The conflict between the clique formation and its destruction leads to $\lamb^-< \lamb^+$.}
    \label{fig:trigrowth}
\end{figure}

A closer inspection of microscopic details explains the sensitivity to the initial state. For an RRG with degree $d$, to remove an edge from a fully connected clique of $d+1$ vertices in the backward rewiring, the system has to destroy $d-1$ triangles simultaneously. Thus, cliques are rarely destroyed during the rewiring process. Conversely, in forward rewiring, constructing a clique of $d+1$ vertices requires successively adding an edge to the existing clique of $d$ vertices. The number of triangles gained from this process starts at $0$ and increases incrementally as one approaches the completion of the clique. Hence, the creation of cliques is a slow process. We demonstrate these two processes in \fig{fig:trigrowth}, using cliques of 3 and 4 vertices as an example. The difficulty of both the creation and destruction of cliques leads to $\lamb^-< \lamb^+$. In practice, it is statistically unlikely that a cluster develops into a highly ordered clique of $d+1$ vertices, yet rewiring any edges away from a cluster is energetically unfavorable, as it can destroy multiple triangles. Thus, clusters with external connections remain as metastable "gas of triangular motifs" collectively forming the inter-cluster subgraph.

\subsection{Edge rewiring in the triangle-poor phase modeled as a chemical reaction under the law of mass action}
\label{sect:3.1}

\begin{figure}[ht]
    \centering
    \includegraphics[width=0.9\linewidth]{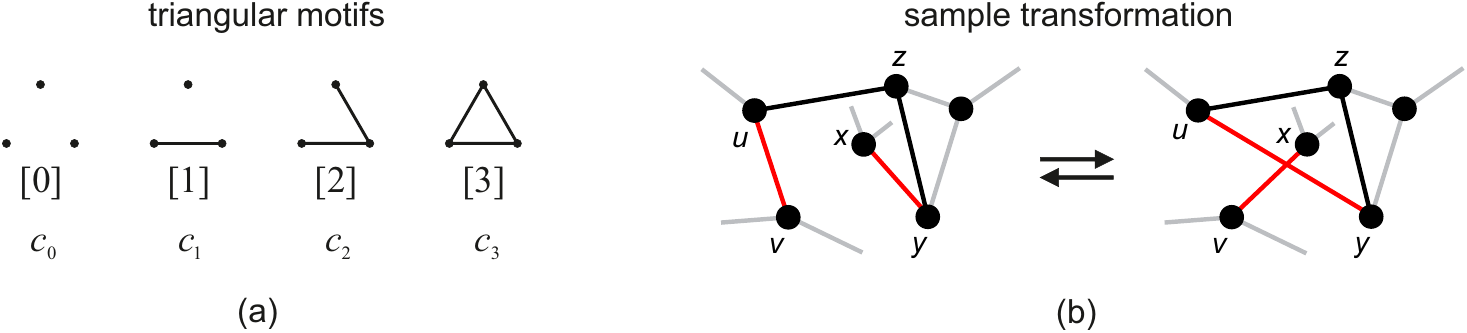}
    \caption{(a) Possible motifs in undirected graphs, $c_0... c_3$ are the concentrations of corresponding motifs; (b) sample of transitions between different types of motifs under a singe rewiring event.}
    \label{fig:triads}
\end{figure}

For small values of $\lamb$, where triangles are formed or destroyed independently, we can describe the kinetics of triangle formations as an elementary "chemical reaction" process following the formalism set out in the paper \cite{massaction} adjusted to the case of rewiring in RRG. The reaction involves four types of 3-point motifs: [0] -- empty motif with no edges, [1] -- motif with one edge, [2] -- motif with two edges, [3] -- motif with three edges, i.e. a complete triangle -- see \fig{fig:triads}a.

Changes in the number of triads of different types at a single rewiring move follow the elementary "chemical reaction" \cite{massaction}
\be
[0]+3[2] \overset{K}{\longleftrightarrow} [3]+3[1]
\label{eq:chem}
\ee
where $K$ is the constant of the chemical reaction \cite{kinetics}. Figure \ref{fig:triads}b provides a sample of particular transformations between motifs satisfying the conservation condition in \eq{eq:chem}. Here, edges $(xy)$ and $(uv)$ are removed, while edges $(xv)$ and $(uy)$ are added. As a result, motif $(xvz)$ passes from type [0] to [1], motifs $(xyz)$ and $(uvz)$ pass from type [2] to [1], and motif $(uyz)$ goes from type [2] to [3]. Overall, 3 motifs of type [1] and 1 of type [3] are generated instead of 3 motifs of type [2] and 1 of type [0].

We denote the concentrations of four types of motifs in the graph by $c_0,...,c_3$, understood as concentration of triads of the corresponding type. The number of triangles, introduced earlier, is recovered by multiplying the concentration of the type [3] motif by the total number of triads $\binom{N}{3}$: $\triangle = \binom{N}{3}\, c_3$. Edge rewiring under vertex degree conservation suggests the existence of the following invariants (valid both for ERG and RRG regardless of the degree constraints)
\be
\sum_{j=0}^{3} c_j = I_1 = 1; \quad \sum_{j=1}^3 j c_j = I_2; \quad c_0+c_3 = I_3
\label{eq:inv}
\ee
where the invariant $I_1$ describes the conservation of all triads, the invariant $I_2$ describes the conservation of all edges, and the invariant $I_3$ follows from the exchange of edges involving solitary full triangles. 

\begin{figure}[ht]
    \centering
    \includegraphics[width=0.6\linewidth]{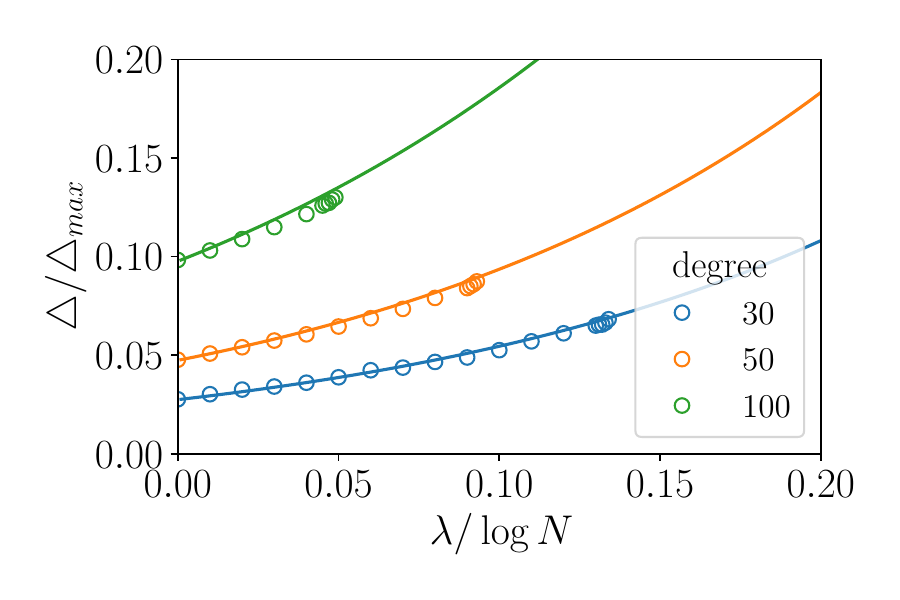}
    \caption{Normalized number of triangles in TPP at $N=1000$ and $d=30,50,100$. Solid lines are the normalized fraction of triangles as described in equation \eqref{eq:tri_ana}.}
    \label{fig:massact}
\end{figure}

Following the proposition of the work \cite{massaction}, we can write the balance equation for the concentration of motifs (triads) participating in the chemical reaction \eq{eq:chem} in in the mean-field approximation when the concentration of triads is small and there is no collective interactions between different rewirings. This approximation in known in chemical kinetics \cite{kinetics} as the "law of mass action" corresponding to the elementary chemical reaction \eq{eq:chem}:
\be
K=e^{\lambda} = \frac{c_3 c_1^3}{c_0 c_2^3}.
\label{eq:mass}
\ee
To obtain the concentration of triangles $c_3$ as a function of $\lamb$, we first rewrite the invariants in terms of $c_1, c_2, c_3$
\be 
\begin{cases}
I_2 = c_1+2c_2+3c_3,  \\ 
I_3 = 1- c_1-c_2.
\end{cases}
\label{eq:inv-mean0}
\ee
This allows us to compute the invariants from the expected values of concentrations $c_1,c_2,c_3$. For a generic RRG at a given degree $d$, these expected values are
\be
\mathbb{E}[c_1] = 3p(1-p)^2; \quad\mathbb{E}[c_2] = 3p^2(1-p); \quad\mathbb{E}[c_3] = p^3,
\label{eq:mean}
\ee
where $p\equiv d/N$. Equations \eq{eq:mean} are exact for large graphs $N\gg 1$. 
We now compute the invariant from the expected values of concentrations
\be 
\begin{cases}
I_2 = \mathbb{E}[c_1]+2\mathbb{E}[c_2]+3\mathbb{E}[c_3], \medskip \\ 
I_3 = 1- \mathbb{E}[c_1]-\mathbb{E}[c_2].
\end{cases}
\label{eq:inv-mean1}
\ee
Substituting the expected value of the concentration into \eqref{eq:inv-mean0}, we obtain the invariants $I_2$ and $I_3$ in terms of $p$:
\be 
\begin{cases}
I_2 = 3p(1-p)^2 +6p^2(1-p)+3p^3, \medskip \\ 
I_3 = 1- 3p(1-p)^2-3p^2(1-p).
\end{cases}
\label{eq:inv-mean}
\ee
The above relations hold for all values of $\lamb$.
As one can see, at the level of equation \eq{eq:inv-mean}, the invariance in the rewiring of RRGs is different from those of the rewiring of ERGs \cite{massaction}. 
Lastly, we express all the concentrations $c_0,c_1,c_2$ in terms of $p$ and $c_3$, using \eqref{eq:inv}, \eqref{eq:inv-mean0} and \eqref{eq:inv-mean}
\be 
\begin{cases}
c_0 =  1 - c_3 - 3p(1-p),\\
c_1 = 3p-6p^2+3c_3, \\ 
c_2 = 3p^2-3c_3.
\end{cases}
\label{eq:conc-p}
\ee
Substituting \eq{eq:conc-p} into \eq{eq:mass}, we arrive at the closed algebraic equation that relates $\lamb$ and $c_3$:
\be
e^{\lambda} = \frac{c_3(p-2p^2+c_3)^3}{(1-c_3-3p(1-p)) (p^2-c_3)^3}.
\label{eq:alg}
\ee
One can straight forwardly verify that at $\lambda = 0$, the solution of \eq{eq:alg} yields the correct value of $c_3$: $c_3(\lambda=0) = p^3$. Since $p=d/N$ and $\bar{c}_3$ are of order $p^3\ll p$, we can make further approximations in equation \eq{eq:alg} for $N\gg 1$. Keeping the leading order in $1/N$ yields the concentration of triangles
\begin{equation}
    c_3 = \frac{d^3e^\lamb}{N^3} + \mcal{O}(N^{-4}).
    \label{eq:alg_approx}
\end{equation}
To compare our derivation with the numerics, we convert the concentration of triangles to the number of triangles by multiplying $c_3$ with $\binom{N}{3}$. However, equation \eq{eq:alg_approx} estimates the number of triangles in RRGs to be $\Delta_{RRG}$ at $\lamb=0$. For numerically generated finite-size RRGs, this is not necessarily true. In order to bypass this finite size effect and accurately assert the validity of the estimation \eq{eq:alg_approx}, we normalize the number of triangles obtained from \eq{eq:alg_approx} such that the number of triangles at $\lamb=0$ from both the numerics and the analytical estimation are equal. Specifically, define
\begin{equation}
\triangle_{an} \equiv \frac{c_3}{\triangle_{max}}\times\frac{\triangle_{num}(\lamb=0)}{c_3(\lamb=0)}
\label{eq:tri_ana}
\end{equation}
where $\triangle_{num}(\lamb=0)$ is the number of triangles obtained numerically at $\lamb=0$. In \fig{fig:massact} the comparison of $\triangle_{an}$ with the numerical simulations for RRGs in the triangle-poor phase for various vertex degrees $d$ is shown. It should be emphasized that no adjustable parameters are used.

Creation of triangles by independent chemical reactions \eq{eq:chem} is limited by the number of available triads of type [2]. In forward rewiring at sufficiently large concentration of full triads there is a bias towards backward reactions in the balance equation \eq{eq:chem} for the fugacity $\lamb$ greater than the transition point $\lamb^+$. This bias implies that the number of available type [2] triads is not sufficient for the rewiring process to arrive at the number of triangles prescribed by $\lamb$ while maintaining chemical equilibrium. Moreover, the prevalence of backward reaction implies that the number of triangles increases through repeated rewiring of the edge, which is already a part of a triangle, from sparser to denser clusters.

\subsection{Criticality of the backward rewiring}
\label{sect:3.2}

In the backward (TRP $\to$ TPP) rewiring, which starts from disconnected cliques, the transition from TRP to TPP occurs at the value $\lamb^-$ at which the free energies of the two corresponding phases coincide. Let us estimate free energies of TRP and TPP independently. This approach is complimentary to the one of \cite{Kochergin:2023prb}.

In TRP the disconnected clique is the unique state with the maximum number of triangles and, hence, with the maximal energy. The free energy of this phase $F_{TRP}$ depends solely on the number of triangles. Thus, 
\be
\frac{F_{TRP}}{k_B T} = -\lamb\triangle_{max} = -\lamb \frac{d(d-1)N}{6}
\ee

In TPP we can estimate the free energy of the graph using the statistics of typical RRGs. At $N\gg 1$ the number of available configurations, $\mcal{N}_{BC}(N,d)$, can be approximated by the Bender-Canfield formula \cite{BC} valid for $\lambda=0$:
\be
\mcal{N}_{BC}(N,d) \approx \sqrt{2}\,e^{-(d^2-1)/4}\left(\frac{(dN)^{d/2}}{e^{d/2}d!}\right)^N.
\label{eq:entropy}
\ee
The number of triangles in TPP is approximately $\triangle_{TPP}\equiv\left(\tfrac{d}{N}\right)^3\,e^{\lambda}$, as follows from \eq{eq:alg_approx}. However, this yields an energy contribution sub-leading in $N$ with respect to \eq{eq:entropy}. Hence, we approximate the free energy in this phase as 
\be
\frac{F_{TPP}}{k_B T} = - \log \mcal{N}_{BC}(N,d) 
\label{eq:free_bender}
\ee

The transition occurs at the point $\lamb_{cr}^-$ defined by the equation 
\be
F_{TRP}(\lambda_{cr}^-)=F_{TPP}
\label{eq:transition}
\ee
Substituting the leading order terms in $N$ for $F_{TRP}(\lambda)$ and $F_{TPP}$ into \eq{eq:transition}, we get
\be
\lamb_{cr}^- = \frac{6}{d(d-1)}\left(\frac{d}{2}\log(dN) - \log (d!) -\frac{d}{2}\right).
\label{eq:lamb_BC}
\ee
We have depicted the estimated critical value, $\lamb_{cr}^-$ by a black vertical line in \fig{fig:hyst}(a)-(c). Note that $\lamb_{cr}^-$ decreases with $d$. The estimate \eq{eq:lamb_BC} becomes more accurate for larger $d$. This estimate correctly explains $\log N$ scaling in the critical value as shown in \fig{fig:hyst}(d). This scaling of the transition point also holds for a rewiring model with Poissonian degree constraints, i.e. Erdos--Renyi graphs, as reported in \cite{grassberger}. 

The attempts to estimate the transition $\lambda^+$ from TPP to TRP in the forward rewiring via the same line of reasoning unfortunately fail. Since graphs in TRP are loosely connected clusters, the corresponding free energy includes, in addition to the energy of dense clusters, the entropic contribution from inter--cluster rewiring which we are not yet able to estimate. However, \fig{fig:hyst}d shows that normalizing $\lamb$ with $\log N$ brings the transition points of RRGs with different $N$ within the same order of magnitude.

In the forward (TPP $\to$ TRP) rewiring, subgraphs with a moderate number of triangles encourage the formation of clusters, since connecting vertices from such subgraphs can generate multiple new triangles. In Section \ref{sect:4}, we show that this is a driving mechanism behind the formation of a scale-free inter-cluster subgraph.

\section{Degree distribution of the inter-cluster subgraph}
\label{sect:4}

In the triangle-rich phase, rewiring a RRG results in a graph that is trapped in a long-living metastable state composed of loosely connected clusters. Below we present results of numerical simulations demonstrating that the degree distribution of the inter-cluster subgraph, defined via the negative Ollivier-Ricci curvature, follows a power law, $P(d) \sim d^{-\gamma}$, with $\gamma \approx 1.5$, over a wide range of graph sizes and degrees. We attribute this universality to the cooperative nature of aggregation of triangles.

We argue that the scale-free behavior of $P(d)$ arises from two key properties of the model: (i) a triangle is a non-local (extended) object, and (ii) forming a new triangle that involves graphs' vertices with a higher clustering coefficient is statistically more favorable than forming an isolated triangle. These features give rise to a driving mechanism similar to preferential attachment. However, unlike the standard Barab\'{a}si--Albert (BA) model \cite{BAmodel}, we do not impose an implicit rule according to which new vertices of a graph are preferentially attached to those with higher vertex degree. Instead, a preferential attachment mechanism in the rewired RRG emerges spontaneously as an intrinsic property of the system. For this reason, it is more appropriate to refer to this mechanism as an "emergent preferential attachment", which shares deep similarities with the mechanism underlying the formation of scale-free citation networks \cite{colman2013complex}, networks grown via likelihood \cite{small2015growing}, and "linear preferential attachment" \cite{Yule1925, Simon1955}.

Below, we propose a degree growth dynamics model that takes into account the emergent preferential attachment mechanism. Our model suggests that the emergent preferential attachment depends on vertex degree superlinearly.

\subsection{Numerics}
\label{sect:4.1}

\begin{figure}[t]
\centering
\includegraphics[width=0.65\textwidth]{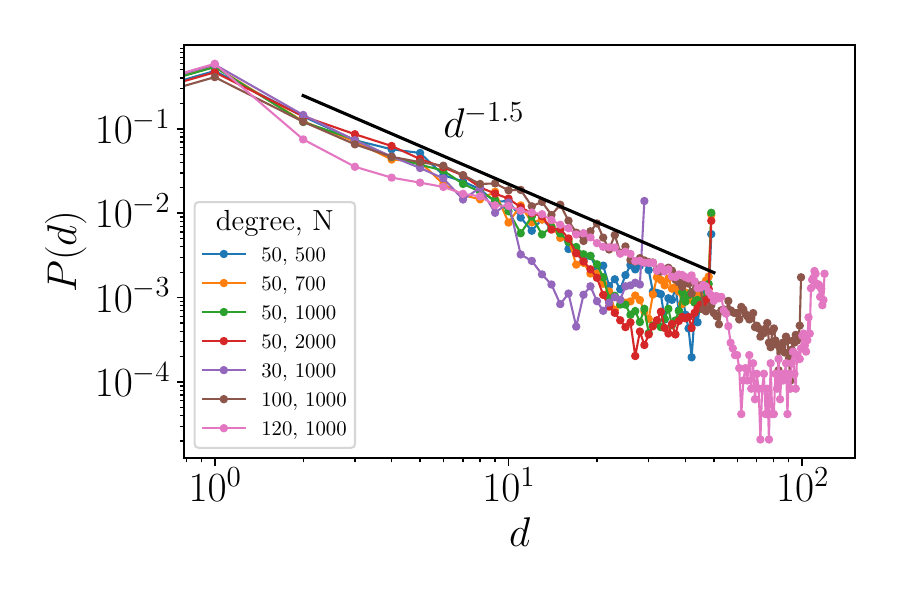}
\caption{Inter-cluster degree distributions of rewired RRG from various degrees and sizes as indicated in the legend. The guide line highlights that all distributions follow power-law $P(d)=d^{-\gamma}$ with exponent $\gamma = 1.50 \pm 0.03$.}
\label{fig:outdegs}
\end{figure}

The inter-cluster degree distribution $P(d)$ of rewired RRGs for different graph sizes and degrees is shown in \fig{fig:outdegs} in the doubly logarithmic scale. Each curve is a histogram constructed from the combined degree statistics of 100 independent inter-cluster subgraphs with the same set of parameters, obtained by forward-rewired RRGs. The fugacity $\lambda$ is set to $2$ (extremely high value) to ensure that the rewired RRGs are surely in the triangle-rich phase. We have verified that larger values of $\lambda$ do not alter the degree distribution. 
 
As shown in \fig{fig:outdegs}, at each set of parameters, the degree distribution $P(d)$ exhibits a power law behavior within a range of degree. This range is enclosed by two regions where the distribution is convex, as most prominently observed in the distribution of rewired RRGs with N=1000, degree = 120. Furthermore, this ranges are shifted to the right the higher the degree of the pre-rewired RRGs is. We report this degree range for the rewired RRGs at each set of parameters in the Table \ref{tab:window}. By performing linear regression on data points within these degree ranges, we estimate the exponent $\gamma$ of the degree distribution to be $\gamma\approx 1.50\pm 0.03$. This value of $\gamma$ significantly deviates from $\gamma_{BA} = 3$ typical for the preferential attraction mechanism in the Barabasi-Albert (BA) model.

\begin{table}[h]
    \centering
    \begin{tabular}{ccc}
         N & degree& degree range \\
         \hline
         500 & 50 & [5,12] \\
700 & 50 & [5,12] \\
1000 & 50 & [5,12] \\
2000 & 50 & [5,12] \\
1000 & 30 & [1,10] \\
1000 & 100 & [8,35] \\
1000 & 120 & [10,40]
    \end{tabular}
    \caption{Degree range of the degree distribution for each graph size and degree, where the data points are taken to estimate the exponent $\gamma$.}
    \label{tab:window}
\end{table}
 
One interesting feature of RRGs in the triangle-rich phase is the distribution of triangle edges across different clusters and inter-cluster subgraphs. A graph in the TRP contains three types of triangles: (i) “point-like” triangles, whose vertices all lie within a single cluster; (ii) “isosceles” triangles, in which two vertices belong to one cluster and the third vertex belongs to the inter-cluster subgraph; and (iii) “equilateral” triangles, whose vertices each lie in the inter-cluster subgraph. According to this classification, almost all vertices are connected to at least one of the clusters. The structure of the inter-cluster subgraph should thus be visualized as groups of isosceles triangles which share a common vertex, and whose bases lie in one or more clusters. These groups are connected by equilateral triangles.

\begin{figure}[ht]
    \centering
    \includegraphics[width=0.65\linewidth]{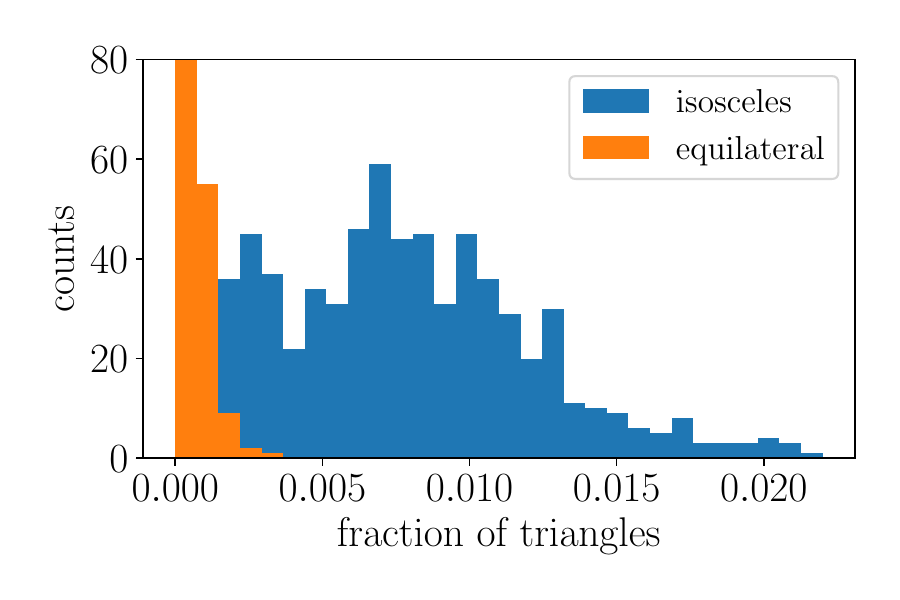}
    \caption{Distribution of isosceles and equilateral triangle as defined in \ref{sect:4} over the same samples of graphs in \fig{fig:outdegs}. For each sample, both numbers of triangle are normalized by its total number of triangles. The distribution of isosceles triangles is much broader than that of the equilateral triangles. There are more isosceles triangles than equilateral triangles in all individual samples.}
    \label{fig:isosceles}
\end{figure}

The creation of an isosceles triangle involves two inter-cluster edges, which originate from two vertices of the same cluster, sharing the third vertex in a different cluster. This process is limited by how likely two inter-cluster edges would share a vertex to close the triangle. Equilateral triangles, each requiring three vertices from different clusters sharing three inter-cluster edges, are less frequent than isosceles triangles, as confirmed by \fig{fig:isosceles}. In particular, the number of isosceles triangles is larger than that of equilateral triangles by an order of magnitude in most samples we have considered.

Isosceles triangles also facilitate the formation of new triangles by providing a shared vertex that can serve as a seed for rewiring edges. The prevalence of isosceles triangles in inter-cluster subgraphs therefore implies an effective attraction toward existing isosceles triangles. These existing seeds are robust, since edges shared by multiple triangles are unlikely to be rewired away, as discussed in previous sections. This mechanism gives rise to an “emergent preferential attachment” in the formation of the inter-cluster subgraphs. Below we propose a mean-field kinetic description of the scale-free inter-cluster subgraph formation observed in the TRP.

\subsection{Mean-field description of spontaneous preferential attachment in clustering of triangles}
\label{sect:4.2}

Here, we provide a possible mechanism behind the apparent power-law degree distribution of the inter-cluster subgraph. We rephrase the increase in isosceles triangles from the rewiring process as a degree growth dynamics with nonlinear preferential attachment, in which a vertex of degree $d$ gains new connections at a rate proportional to $d^\alpha$. This process results asymptotically in a power-law vertex degree distribution.

We start from the intermediate stage of rewiring, where the system consists of partially formed, unsaturated clusters and an inter-cluster subgraph. Each vertex either belongs to a cluster or to the inter-cluster subgraph, while edges can still be rewired into clusters. The inter-cluster subgraph is assumed to be random and not yet possessing a scale-free structure. The number of triangles increases due to edge rewiring that completes a new triangle either (i) within a cluster, or (ii) within the inter-cluster subgraph. Since clusters are denser than the inter-cluster subgraph, a rewired edge is more likely to close a triangle through process (i) than through process (ii). However, it is the triangle formation in process (ii) that leads to the emergence of the scale-free structure in the inter-cluster subgraph. We interpret process (i) as the “fast” removal of edges from the inter-cluster subgraph, while process (ii) corresponds to the formation of a scale-free network within the inter-cluster subgraph through the creation of isosceles triangles, as discussed in section \ref{sect:4.1}. It is precisely this process that is discussed below.

Each creation of isosceles triangles in one rewiring step consists of closing a triad. As the base of the newly formed triangle belongs to a cluster, creating isosceles triangle is equivalent to growing the inter-cluster subgraph by attaching a new vertex to the already existing one. Further more, vertices that are shared among more triangles survive more rewiring attempts, and are more likely to gain more edges. We assume that this supplies the growth of inter-cluster subgraph with "nonlinear preferential attachment" in degree, meaning that the vertices of degree $d$ gain more degrees at the rate $d^\alpha$. This preferential attachment mechanism 
spontaneously emerges from collective interaction of extended objects -- triangles.

We can describe this growth process within a mean-field framework. We adopt a kinetic approach to network evolution in the spirit of \cite{ben2004kinetic}. Let $K_j(t)$ denote the number of inter-cluster vertices of degree $j$, and let $t = \sum_j K_j$ be the total number of vertices in the inter-cluster subgraph. Within the mean-field description, the evolution of the inter-cluster subgraph (and hence the dynamics of vertex degrees) is governed exclusively by the formation of isosceles triangles. Under the assumption of nonlinear preferential attachment, the degree of a vertex changes at a rate proportional to some power $\alpha>1$ of its current degree. In particular, vertices of degree $j$ evolve into vertices of degree $(j+1)$, so the number of degree-$j$ vertices decreases at rate $j^\alpha K_j / t$. At the same time, vertices of degree $(j-1)$ evolve into vertices of degree $j$, leading to an increase in the number of degree-$j$ vertices at rate $(j-1)^\alpha K_{j-1}/t$.

Combining these loss and gain terms, we obtain the evolution equation governing the number of vertices of degree $j$:
\be
\frac{dK_j}{dt} = \frac{1}{t}\left[(j-1)^\alpha K_{j-1} - j^\alpha K_j + \delta_{1,j}\right]
\label{eq:lPA}
\ee
The Kronecker delta is added to take into account vertex creation so that the rate of $K_1$ is non-negative
\be
\frac{dK_1}{dt} = \frac{1}{t}\left(1- K_1\right)
\ee
Equation \eqref{eq:lPA} admits the ansatz $K_j(t) = c_j\,t$ where the linearity in $t$ reflects the linear preferential attachment mechanism and the frequency $c_j$ of each vertex degree is independent of $t$. Substituting the linear ansatz $K_j(t) = c_j\,t$ in \eqref{eq:lPA} we get a time-independent recursion for $c_j$
\be
\frac{c_j}{c_{j-1}} = \frac{(j-1)^\alpha}{j^\alpha+1} \underset{j\gg1}{\sim} 1-\frac{\alpha}{j}.
\label{eq:cjcj1}
\ee
Asymptotically, for large $j$, \eq{eq:cjcj1} is consistent with $c_j \propto j^{-\alpha}$ providing a power-law distribution with the critical exponent $\gamma=\alpha$. According to the numerics of Section \ref{sect:4.1}, the emergent preferential attachment from triangle motif yields $\alpha \approx 1.5$.

For $\alpha = 1$, Eq. \eqref{eq:cjcj1} behaves asymptotically as $c_j/c_{j+1} \sim (1-2/j)$ for large $j$, leading to a power law degree distribution with $\gamma=2$. This reproduces the degree distribution of classical models with asymmetric reinforcement, such as the Simon-Yule process \cite{Simon1955,Yule1925} and directed preferential attachment models \cite{Krapivsky2000, KrapivskyRedner2001} (see also \cite{PArev} for a review of other systems with power law distribution). In these systems, attachment events reinforce only one side of a connection, resulting universally in power-law degree distributions with the aforementioned exponent $\gamma=2$. The formation of inter-cluster subgraph via the creation of isosceles triangle is also one-sided-type, since the emergent preferential attachment affects the shared vertex, but not the bases of triangles. However, the exponent $\alpha>1$ suggests the nonlinearity of this emergent effect.

\section{Discussion}
\label{sect:5}

Here we summarize the main results obtained in this work. Our analysis of the emergence of scale-free inter-cluster subgraphs with the exponent $\gamma \approx 1.5$ would be incomplete without discussing how such scale-free networks are embedded into hyperbolic space, as well as their similarities to and distinctions from the widely-known construction proposed in \cite{Boguna1,Krioukov,Boguna2}. 

\subsection{Results}

Triangle-driven rewiring of random regular graphs (RRGs) exhibits the following main features. For small values of the fugacity of closed triangles, $\lambda$, the number of triangles is proportional to $e^{\lambda}$ and follows the law of mass action described by the “chemical reaction” among different types of triads. As $\lambda$ increases, a first-order transition occurs at the value $\lambda^{\pm}$ that depends on the direction of rewiring: forward, or backward.

Forward rewiring at $\lamb>\lamb^+$, leads to a long-lived metastable state consisting of loosely connected clusters. The inter-cluster subgraph is scale-free, with exponent $\gamma \approx 1.5$, independent of the graph degree and size. By contrast, backward rewiring, starting from disconnected cliques (the state with the maximal number of triangles in an RRG), yields a transition at $\lambda^{-} < \lambda^{+}$. In both cases, the critical transition values, $\lambda^{\pm}$, scale as $\log N$.

The first order transition in the backward rewiring occurs between two phases with distinct contributions to the free energy: in one phase the main contribution comes from the entropy which scales as $N\log N$, while in the other phase -- from the number of triangles, which scales as $N$. This results in the peculiar $\sim \log N$ scaling of the critical triangle fugacity $\lamb^-$. Numerical simulations show that this scaling is valid also for $\lamb^+$.

The scale-free property of the inter-cluster subgraph arises from an emergent nonlinear preferential attachment driven by an effective "attraction" of triangles sharing the same vertices. The mechanism behind a scale-free graph formation, which can be denoted as a “spontaneous preferential attachment”, deals with the creation of a new triangle attached to a group of already existing triangles with a high clustering coefficient. Such an attachment is statistically more favorable than the creation of a new solitary triangle. This mechanism has much in common with the asymmetric reinforcement \cite{Simon1955,Yule1925}, however its nonlinearity results in degree distribution with exponent $\gamma=1.5$ instead of more common result $\gamma=2$.  

Similar motif-enrichment rewiring of RRGs has been shown to induce changes in global graph properties. In \cite{Gorsky:2021prd}, the spectral properties of the adjacency matrix were also found to change under rewiring of RRGs that increases the number of squares (closed loops of length four). We encourage further investigation of the relationship between motif enrichment and spectral properties of the network.

In our work, we do not merge vertices within the same cluster into a single "composite" vertex as in the previous study \cite{pandemic}, instead, we treat vertices inside a cluster as individual entities and extract only the edges connecting different clusters, while disregarding connections with edges internal to the same cluster. Modifying this protocol by grouping vertices within each cluster into a composite vertex could alter the exponent $\gamma$, potentially making $\gamma$ degree-dependent. Further investigation shall provide insight into possible vertex grouping that, when applied to an existing inter-cluster subgraph, could serve as a tool of tuning the exponent $\gamma$ in the scale-free distribution $P(d)$.

The prevalence of scale-free networks in real-world systems has recently been challenged by data-driven studies \cite{Broido_Clauset,Caldarelli}. Our work demonstrates that a power-law degree distribution may be hidden within a subgraph of the full network. As shown in a series of previous works \cite{tunneling, pandemic, new-localization, massaction, Gorsky:2020cnaa008, Gorsky:2021prd, Kochergin:2023prb}, a power-law degree distribution within a subgraph can significantly influence the global properties of the entire network, causing it to exhibit features typically associated with scale-free networks. This suggests that the classification of scale-free networks may need to be extended beyond the presence of a global power-law degree distribution.

\subsection{Scale-free graphs from hyperbolic embedding}
\label{sect:5.1}

Let us remind the main steps of embedding scale-free graphs into the hyperbolic manifold of a constant negative curvature $-1$ suggested in the pioneering works \cite{Boguna1,Krioukov,Boguna2}. Instead of the Poincar\'e disc, we consider the Poincar\'e upper half-plane $\mathbb{H}^2 = \{(x,y):y>0\}$ with the metric $ds^2$ defined in the standard way, $ds^2 = \tfrac{1}{y^2}(dx^2 + dy^2)$, where $(x,y)\in\mathbb{R} \times \mathbb{R}_{>0}$. Using the field-theoretic language, consider the Dirichlet problem for a massive scalar field $\phi(x,y)$ which has a point source at some cutoff distance $y=y_0=e^{-d}$. The bulk "equation of motion" for the field $\phi(x,y)$ is determined by the Laplace equation
\be
(\Delta_{\mathbb{H}^2} - m^2)\phi = 0
\ee
where $\Delta_{\mathbb{H}^2} = y^2(\partial_x^2 + \partial_y^2)$ is the Laplace-Beltrami operator on $\mathbb{H}^2$ and $m$ is the mass of the scalar field. To simplify the description and highlight the main statements, let us begin with the case $m=0$. The solution of the massless theory in the upper half-plane $y > y_0$ with initial point $(x_0,y_0)$ is the Poisson kernel:
\be
\phi(x,y) = \frac{1}{\pi}\frac{y-y_0}{(x-x_0)^2+(y-y_0)^2}, \quad \int_{-\infty}^{\infty} \phi(x,y)dx = 1
\ee
For a point at the same horizontal coordinate as the source ($x = x_0$) one can write
\be
\phi(y) = \frac{1}{\pi(y - y_0)}
\ee
The hyperbolic distance $\mu$ between $(x_0, y)$ and the source at $(x_0, y_0)$ is
\be
\mu = \int_{y_0}^{y} \frac{dy'}{y'} = \ln \frac{y}{y_0} \quad \Rightarrow \quad y = y_0 e^{\mu}
\label{eq:mu}
\ee
For large $\mu$ one has $y \gg y_0$, and we can express the field $\phi$ through the geodesic distance
\be
\phi \approx \frac{1}{\pi y} = \frac{1}{\pi y_0}e^{-\mu} 
\label{eq:phi-mu}
\ee

Consider now graphs embedded into the Poincar\'e upper half-plane $\mathbb{H}^2$. Following \cite{Krioukov}, let us introduce a generalized "hidden variable", representing the node "popularity" or expected degree, parameterized by a scaling exponent $\beta > 0$:
\be
\varkappa = e^{\beta\mu}
\label{eq:kappa-beta}
\ee
Using \eq{eq:mu}, we can express $\varkappa$ directly in terms of the vertical coordinate $y$ and the cutoff $y_0$:
\be
\varkappa = \left(\frac{y}{y_0}\right)^{\beta}
\ee
Suppose that nodes are distributed uniformly over the hyperbolic plane. In the upper half-plane this implies using the invariant area measure $dA = \tfrac{dx\,dy}{y^2}$. For a fixed horizontal coordinate $x$, the marginal distribution of $y$ is
\be
\rho(y)\, dy \propto \frac{dy}{y^2}
\ee
Inverting the relation $\varkappa = (y/y_0)^{\beta}$, we express $y$ and its differential $dy$ in terms of $\varkappa$:
\be
y = y_0 \varkappa^{1/\beta}, \qquad dy = \frac{y_0}{\beta} \varkappa^{\frac{1}{\beta} - 1} d\varkappa
\ee
Substituting these expressions into the distribution $\rho(y)\, dy$, we obtain the probability density of the expected degree $\varkappa$:
\be
\rho(\varkappa)\, d\varkappa \propto \frac{1}{\beta y_0} \varkappa^{-\left(1 + \frac{1}{\beta}\right)} d\varkappa
\label{eq:rho-kappa}
\ee
Thus, $\rho(\varkappa) \sim \varkappa^{-\gamma}$, meaning that the vertex degree distribution exhibits a power-law tail with a $\beta$-dependent critical exponent
\be
\gamma = 1 + \frac{1}{\beta}
\label{eq:gamma-beta}
\ee

It is important to emphasize that the difference in the exponent $\gamma$ across various hyperbolic network models is not connected with the choice of coordinate system. As the Poincar\'e upper half-plane $\mathbb{H}^2$ and the Poincar\'e disc are globally isometric, all such models assume a uniform node density with respect to the invariant hyperbolic area measure $dA$. Rather, the value of $\beta$ (and consequently the scale-free exponent $\gamma$) arises entirely from the physical rules mapping the expected degree $\varkappa$ onto the geodesic distance $\mu$ (or equivalently, onto the bulk propagator $\phi \sim e^{-\mu}$ via $\varkappa \sim \phi^{-\beta}$):
\begin{enumerate}

\item Standard geometric "disc-intersection" rule ($\beta = \tfrac{1}{2}$ and $\gamma = 3$): In \cite{Krioukov}, two vertices are linked if their hyperbolic distance lies within a global cutoff radius $R$ (in the zero-temperature limit at constant curvature $-1$). Due to the hyperbolic triangle inequality, the effective connecting area of a hyperbolic disc at radial distance $\mu$ from the origin scales with the \emph{square root} of the exponential distance: $\varkappa \sim e^{\mu/2}$, corresponding to $\beta = \tfrac{1}{2}$. Substituting $\beta = \tfrac{1}{2}$ into \eq{eq:gamma-beta} immediately recovers the standard Barab\'asi--Albert exponent $\gamma = 3$ \cite{BAmodel}.

\item Linear propagator scaling ($\beta = 1$ and $\gamma = 2$): In our construction, connectivity among the inter-cluster subgraphs is governed not by the overlapping area of two hyperbolic discs, but rather by the bulk Green function (Poisson kernel), $\phi \sim e^{-\mu}$, of the Laplace-Beltrami operator. Setting $\varkappa \sim \phi^{-1} = e^{\mu}$ corresponds to $\beta = 1$, which provides a \emph{linear} scaling in $e^{\mu} = y/y_0$. Here, the variable $y$ effectively measures the "hierarchical levels": $y \to y_0$ corresponds to the peripheral clusters, while larger $y$ corresponds to central inter-cluster hubs. Substituting $\beta = 1$ into \eq{eq:gamma-beta} yields the scale-free exponent $\gamma = 2$. This geometric mechanism is physically consistent with linear preferential attachment and, in particular, with the classical Simon--Yule evolutionary process \cite{Simon1955,Yule1925}. Direct linear proportionality ($\varkappa = y/y_0 = e^{\mu}$) implies that the capacity of a vertex to attract incoming links scales linearly with its height (hierarchical level) $y$.

\item Massive bulk field generalization ($\beta = \beta(m)$): More generally, for a massive scalar field ($m > 0$) in $\mathbb{H}^2$, the asymptotic decay of the bulk Green function scales as $\phi \sim e^{-\delta \mu}$, where $\delta = \tfrac{1}{2} + \sqrt{\tfrac{1}{4} + m^2}$ is the conformal scaling dimension. Setting $\varkappa \sim \phi^{-1}$ naturally yields exponents $\beta = \delta \ge 1$, continuously tuning the scale-free exponent across the range $1 < \gamma \le 2$.

In Appendix \ref{sect:app1}, we discuss that the Ollivier--Ricci curvature of the scale-free subgraph is not strictly constant, but is varying across edges of the graph. We conjecture that spatial variation in negative curvature modifies the dependence of node popularity on geodesic distance $\mu$, thereby shifting the effective exponent $\beta$ in $\varkappa = e^{\beta\mu}$ since fluctuations around a negatively curved background can act as an effective mass term or potential disorder for the bulk field, providing $m > 0$ and, hence, $\beta > 1$.

\end{enumerate}

Interesting feature of triangle-reach phase consists in prevalence of isosceles triangles which also facilitates the formation of new triangles by providing a shared vertex that can serve as a seed for rewiring edges. The prevalence of isosceles triangles in inter-cluster subgraphs implies an effective attraction toward existing isosceles triangles. We permit ourselves to make a debatable conjecture that the inter-cluster subgraph in the triangle-rich phase should not be viewed merely as a kinetic artifact of incomplete clustering. Rather, it constitutes a distinct topological regime in which a spanning inter-cluster subgraph is stabilized against complete defragmentation into a disconnected set of cliques, by a sparse, structurally rigid gas of apex-shared isosceles triads.

In following works it would be interesting to explore possible roles of these isosceles triangles in the stabilization of the triangle-rich phase (TPP) from the point of view of "Efimov effect" \cite{akkermans, hyp-efimov1, hyp-efimov2}. In the Efimov effect, two particles fail to form a stable bound state on their own, yet a third particle mediates an effective $1/y^2$ attraction that forces all three particles to form a long-living trimer. Analogously, in our RRG rewiring dynamics, an isolated inter-cluster edge connecting one cluster to another is entropically unstable and is rapidly rewired away.
Hence, two vertices from two different clusters do not form a stable pair. However, two vertices in the same cluster and the third vertex from another cluster form an isosceles triangle, which is topologically protected against rewiring.

\begin{acknowledgments}

We are grateful to Oleg Evnin, Alexander Gorsky, Olga Valba, and Andrey Zelenskiy for valuable discussions.

\end{acknowledgments}

\begin{appendix}

\section{Reminder on Ollivier-Ricci curvature and optimal transport on a graph}
\label{sect:app1}

\subsection{Definitions}

Consider two mass distributions, $m_x$ and $m_y$, defined on the vertices: one centered at vertex $x$, and another -- at vertex $y$. The Wasserstein distance on graphs $W(m_x,m_y)$ is the minimum total travel distance (minimum "cost") to convert the distribution $m_x$ into $m_y$. Over the set of vertices $V$, one defines a "transport plan" $\pi(u,v): V\times V \rightarrow [0,1]$, which is the amount of mass $m_x(v)$ at $v$ to be moved to $u$. Since $m_x(v)$ has to be moved from vertex $v$, and, on the other hand, $m_y(v)$ has to be received on vertex $v$, one has a set of constraints $\sum_{v'\in V} \pi(u,v') = m_x(u)$ and $\sum_{u'\in V} \pi(u',v) = m_y(v)$. The Wasserstein distance is thus defined as follows:
\be
W(m_x,m_y) = \underset{\pi}{\text{inf}}\left[\sum_{u,v\in V}\pi(u,v)d(u,v)\right]
\ee
with the constraints above, where $d(u,v)$ is the length of the shortest path $(u,v)$. The Ollivier-Ricci curvature (ORC) of edge $(x,y)$ is  
\be
\kappa_{x,y} = 1-\frac{W(m_x,m_y)}{d(x,y)}.
\ee
Following \cite{comm_detect}, the probability mass distribution is 
\be
m_x^a(x_i) = \begin{cases} \disp a &\text{if }x_i =x, \\ \disp \frac{1-a}{C}e^{-d^2(x,x_i)} & \disp x_i \text{ is connected to } x, \\ \disp 0 & \text{otherwise}
\end{cases}
\ee
The so-called "idleness" parameter $a\in[0,1]$ tunes how sharp the mass distribution is, with less sharp distribution leading to higher contrast in the curvature. We find that its value has little effect on the cluster detection, as reported in \cite{alpha}. In this work, we choose $a=0$ for the best contrast of the positive and negative curvature. Our model concerns unweighted graphs where every vertex is at distance 1 from its neighbor, so each neighbor of $x$ receives the same mass $1/d_x$ with $d_x$ being the degree of vertex $x$.

\subsection{Curvature and triangles}

For an edge $(x,y)$ of a $d$-regular graph with $\Delta$ triangles, Jost and Liu \cite{JostLiu2014} proved the inequality
\be
-2+\frac{4}{d}+\frac{3\Delta}{d}\le\kappa_{x,y}\le\frac{\Delta}{d} \qquad (\Delta\le d-2).
\label{eq:jl}
\ee
where $\kappa_{x,y}$ is the ORC of the exge $(x,y)$. Each triangle raises the lower bound by $3/d$. An edge with no short cycles has the tree-like value $-2+4/d$. For an edge of a fully connected cluster of $d+1$ vertices, $\Delta=d-1$ and the curvature equals $(d-1)/d$, which is the value for the complete graph \cite{Lin2011, Srinivasan2025}. Cycles of length 4 and 5 through the edge raise the curvature above the lower bound, while longer cycles do not affect it \cite{Bhattacharya2015}.

\begin{figure}[ht]
    \centering
    \includegraphics[width=\linewidth]{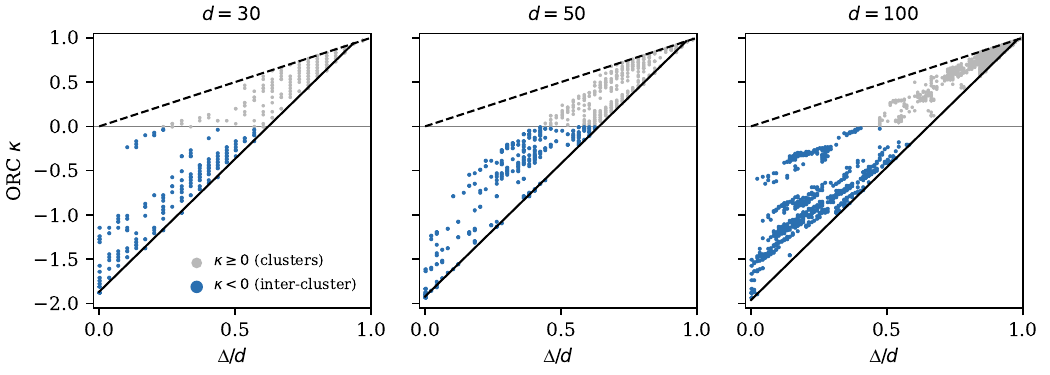}
    \caption{ORC of the edges of rewired RRGs with $N=1000$ and $\lambda=2$, for $d=30$, 50 and 100 (one graph each), against the number of triangles $\Delta$ on the edge divided by $d$. Gray: edges with $\kappa\ge 0$ (clusters); blue: edges with $\kappa<0$ (inter-cluster subgraph). Solid line: lower bound $-2+4/d+3\Delta/d$; dashed line: upper bound $\Delta/d$ of \eq{eq:jl} \cite{JostLiu2014}.}
    \label{fig:orc_bound}
\end{figure}

Figure \ref{fig:orc_bound} shows the ORC of every edge against $\Delta/d$. For every edge ORC lies between the two bounds of \eq{eq:jl}. Edges of complete clusters have ORC $\kappa=(d-1)/d$ exactly, and for $d=30$ and $50$ the most negative curvature equals the tree-like value $-2+4/d$.

\subsection{Clusters and the inter-cluster subgraph}

Edges inside clusters carry many triangles and have curvature close to $(d-1)/d$, while the tree-like value is negative. The sign of the ORC therefore separates edges inside clusters from bottleneck edges between them, as in ORC-based community detection \cite{comm_detect, sia2019ollivier, Srinivasan2025}. As defined in Section \ref{sect:2}, clusters are formed by the edges with positive ORC and their vertices, and the inter-cluster subgraph -- by the connected set of edges with negative ORC and their vertices (gray and blue points in \fig{fig:orc_bound}, respectively). Each edge belongs either to a cluster or to the inter-cluster subgraph, while a vertex can belong to both. The degree distribution of Section \ref{sect:4} counts the edges with negative ORC only.

The ORC of the inter-cluster subgraph is not constant, as noted in Section \ref{sect:5}: it varies from edge to edge between the tree-like value and zero, according to the number of triangles and of 4- and 5-cycles through each edge -- as shown in \fig{fig:orc_bound}. Edges whose endpoints share few neighbors lie near the lower bound, while edges whose endpoints share more neighbors approach $\kappa=0$.

\end{appendix}

\bibliographystyle{unsrt}
\bibliography{bib}

@article{PastorSatorras2001,
  author = {Pastor-Satorras, Romualdo and Vespignani, Alessandro},
  title = {Epidemic spreading in scale-free networks},
  journal = {Physical Review Letters},
  volume = {86},
  number = {14},
  pages = {3200--3203},
  year = {2001},
  publisher = {American Physical Society},
  doi = {10.1103/PhysRevLett.86.3200}
}

@article{Lee2005,
  author = {Lee, D. S.},
  title = {Synchronization transition in scale-free networks: Clusters of synchrony},
  journal = {Physical Review E},
  volume = {72},
  number = {2},
  pages = {026208},
  year = {2005},
  publisher = {American Physical Society},
  doi = {10.1103/PhysRevE.72.026208}
}

@article{Song2005,
  author = {Song, Chaoming and Havlin, Shlomo and Makse, Hern{\'a}n A.},
  title = {Self-similarity of complex networks},
  journal = {Nature},
  volume = {433},
  number = {7024},
  pages = {392--395},
  year = {2005},
  publisher = {Nature Publishing Group},
  doi = {10.1038/nature03248}
}

@article{Chen2017,
  author = {Chen, X. S. and Ding, Y. M. and Meng, J. and Fan, J. F. and Ye, F. F.},
  title = {Statistical properties of random clique networks},
  journal = {Frontiers of Physics},
  volume = {12},
  number = {5},
  pages = {128909},
  year = {2017},
  publisher = {Springer},
  doi = {10.1007/s11467-017-0682-x}
}

@article{PMC3151270,
  author = {Bialonski, S. and Lehnertz, K.},
  title = {Unraveling spurious properties of interaction networks with tailored random networks},
  journal = {PLoS ONE},
  volume = {6},
  number = {8},
  pages = {e22826},
  year = {2011},
  publisher = {Public Library of Science},
  doi = {10.1371/journal.pone.0022826}
}

@article{Grassberger,
  title = {Communities, clustering phase transitions, and hysteresis: Pitfalls in constructing network ensembles},
  author = {Foster, David and Foster, Jacob and Paczuski, Maya and Grassberger, Peter},
  journal = {Phys. Rev. E},
  volume = {81},
  issue = {4},
  pages = {046115},
  numpages = {12},
  year = {2010},
  month = {Apr},
  publisher = {American Physical Society},
  doi = {10.1103/PhysRevE.81.046115},
  url = {https://link.aps.org/doi/10.1103/PhysRevE.81.046115}
}

@article{tunneling,
  title = {Eigenvalue tunneling and decay of quenched random network},
  author = {Avetisov, V. and Hovhannisyan, M. and Gorsky, A. and Nechaev, S. and Tamm, M. and Valba, O.},
  journal = {Phys. Rev. E},
  volume = {94},
  issue = {6},
  pages = {062313},
  numpages = {6},
  year = {2016},
  month = {Dec},
  publisher = {American Physical Society},
  doi = {10.1103/PhysRevE.94.062313},
  url = {https://link.aps.org/doi/10.1103/PhysRevE.94.062313}
}

@article{pandemic,
  title = {Self-isolation or borders closing: What prevents the spread of the epidemic better?},
  author = {Valba, O. and Avetisov, V. and Gorsky, A. and Nechaev, S.},
  journal = {Phys. Rev. E},
  volume = {102},
  issue = {1},
  pages = {010401},
  numpages = {6},
  year = {2020},
  month = {Jul},
  publisher = {American Physical Society},
  doi = {10.1103/PhysRevE.102.010401},
  url = {https://link.aps.org/doi/10.1103/PhysRevE.102.010401}
}

@article{Strauss,
author = {Strauss, David},
title = {On a General Class of Models for Interaction},
journal = {SIAM Review},
volume = {28},
number = {4},
pages = {513-527},
year = {1986},
doi = {10.1137/1028156},
URL = {https://doi.org/10.1137/1028156},
eprint = {https://doi.org/10.1137/1028156}
}

@article{burda,
  title = {Network transitivity and matrix models},
  author = {Burda, Z. and Jurkiewicz, J. and Krzywicki, A.},
  journal = {Phys. Rev. E},
  volume = {69},
  issue = {2},
  pages = {026106},
  numpages = {10},
  year = {2004},
  month = {Feb},
  publisher = {American Physical Society},
  doi = {10.1103/PhysRevE.69.026106},
  url = {https://link.aps.org/doi/10.1103/PhysRevE.69.026106}
}

@article{Newman,
  title = {Solution for the properties of a clustered network},
  author = {Park, Juyong and Newman, M. E. J.},
  journal = {Phys. Rev. E},
  volume = {72},
  issue = {2},
  pages = {026136},
  numpages = {5},
  year = {2005},
  month = {Aug},
  publisher = {American Physical Society},
  doi = {10.1103/PhysRevE.72.026136},
  url = {https://link.aps.org/doi/10.1103/PhysRevE.72.026136}
}

@article{comm_detect, 
title={Community Detection on Networks with Ricci Flow}, 
volume={9}, 
DOI={10.1038/s41598-019-46380-9}, 
number={1}, 
journal={Scientific Reports}, 
publisher={Springer Science and Business Media LLC}, 
author={Ni, Chien-Chun and Lin, Yu-Yao and Luo, Feng and Gao, Jie},
year={2019}
}

@article{ollivier1,
title = {Ricci curvature of Markov chains on metric spaces},
journal = {Journal of Functional Analysis},
volume = {256},
number = {3},
pages = {810-864},
year = {2009},
issn = {0022-1236},
doi = {https://doi.org/10.1016/j.jfa.2008.11.001},
url = {https://www.sciencedirect.com/science/article/pii/S002212360800493X},
author = {Yann Ollivier},
}

@article{Lin2011,
  author    = {Yong Lin and Linyuan Lu and Shing-Tung Yau},
  title     = {Ricci curvature of graphs},
  journal   = {Tohoku Mathematical Journal},
  volume    = {63},
  number    = {4},
  pages     = {605--627},
  year      = {2011},
  doi       = {10.2748/tmj/1325886283},
  note      = {Adapts Ollivier's Wasserstein-based curvature to graph theory}
}

@article{WattsStrogatz1998,
  author    = {Duncan J. Watts and Steven H. Strogatz},
  title     = {Collective dynamics of 'small-world' networks},
  journal   = {Nature},
  volume    = {393},
  number    = {6684},
  pages     = {440--442},
  year      = {1998},
  doi       = {10.1038/30918}
}

@article{BC,
title = {The asymptotic number of labeled graphs with given degree sequences},
journal = {Journal of Combinatorial Theory, Series A},
volume = {24},
number = {3},
pages = {296-307},
year = {1978},
issn = {0097-3165},
doi = {https://doi.org/10.1016/0097-3165(78)90059-6},
url = {https://www.sciencedirect.com/science/article/pii/0097316578900596},
author = {Edward A Bender and E.Rodney Canfield}
}

@INPROCEEDINGS{alpha,
  author={Ni, Chien-Chun and Lin, Yu-Yao and Gao, Jie and David Gu, Xianfeng and Saucan, Emil},
  booktitle={2015 IEEE Conference on Computer Communications (INFOCOM)}, 
  title={Ricci curvature of the Internet topology}, 
  year={2015},
  volume={},
  number={},
  pages={2758-2766},
  doi={10.1109/INFOCOM.2015.7218668}}

@article{birth_geo,
title={The birth of geometry in exponential random graphs}, 
volume={54}, 
DOI={10.1088/1751-8121/ac2474}, 
number={42}, 
journal={Journal of Physics A: Mathematical and Theoretical}, 
publisher={IOP Publishing}, 
author={Akara-pipattana, Pawat and Chotibut, Thiparat and Evnin, Oleg}, 
year={2021}, 
pages={425001}
}

@article{massaction,
  title = {Islands of Stability in Motif Distributions of Random Networks},
  author = {Tamm, M. V. and Shkarin, A. B. and Avetisov, V. A. and Valba, O. V. and Nechaev, S. K.},
  journal = {Phys. Rev. Lett.},
  volume = {113},
  issue = {9},
  pages = {095701},
  numpages = {5},
  year = {2014},
  month = {Aug},
  publisher = {American Physical Society},
  doi = {10.1103/PhysRevLett.113.095701},
  url = {https://link.aps.org/doi/10.1103/PhysRevLett.113.095701}
}

@article{Simon1955,
  author = {Simon, H. A.},
  title = {On a Class of Skew Distribution Functions},
  journal = {Biometrika},
  volume = {42},
  pages = {425--440},
  year = {1955}
}

@article{Yule1925,
  author = {Yule, G. U.},
  title = {A Mathematical Theory of Evolution},
  journal = {Philosophical Transactions of the Royal Society B},
  volume = {213},
  pages = {21--87},
  year = {1925}
}

@article{colman2013complex,
  title = {Complex scale-free networks with tunable power-law exponent and clustering},
  author = {Colman, E. R. and Rodgers, G. J.},
  journal = {Physica A: Statistical Mechanics and its Applications},
  volume = {392},
  number = {21},
  pages = {5501--5510},
  year = {2013},
  publisher = {Elsevier},
  doi = {10.1016/j.physa.2013.06.063},
  url = {https://arxiv.org/abs/1307.7389},
  archiveprefix = {arXiv},
  eprint = {1307.7389},
  primaryclass = {physics.soc-ph},
}

@article{Krapivsky2000,
  author = {Krapivsky, P. L. and Redner, S. and Leyvraz, F.},
  title = {Connectivity of Growing Random Networks},
  journal = {Phys. Rev. Lett.},
  volume = {85},
  pages = {4629},
  year = {2000}
}

@article{KrapivskyRedner2001,
  author = {Krapivsky, P. L. and Redner, S.},
  title = {Organization of Growing Random Networks},
  journal = {Phys. Rev. E},
  volume = {63},
  pages = {066123},
  year = {2001}
}

@article{ben2004kinetic,
  title = {Kinetic theory of random graphs: From paths to cycles},
  author = {Ben-Naim, E. and Krapivsky, P. L.},
  journal = {Physical Review E},
  volume = {71},
  number = {2},
  pages = {026129},
  year = {2005},
}

@article{Newman2005,
  author = {Newman, M. E. J.},
  title = {Power laws, Pareto distributions and Zipf’s law},
  journal = {Contemporary Physics},
  volume = {46},
  pages = {323--351},
  year = {2005}
}

@article{small2015growing,
  title = {Growing optimal scale-free networks via likelihood},
  author = {Small, Michael and Li, Yingying and Stemler, Thomas and Judd, Kevin},
  journal = {Physical Review E},
  volume = {91},
  number = {4},
  pages = {042801},
  year = {2015}
}

@article{Boguna1,
  title     = {Curvature and temperature of complex networks},
  author    = {Bogu{\~n}{\'a}, Mari{\'a}n and Papadopoulos, Fragkiskos and Krioukov, Dmitri},
  journal   = {Physical Review E},
  volume    = {80},
  number    = {3},
  pages     = {035101(R)},
  year      = {2009},
  doi       = {10.1103/PhysRevE.80.035101},
  url       = {https://doi.org/10.1103/PhysRevE.80.035101},
  publisher = {American Physical Society}
}

@article{Krioukov,
  title     = {Hyperbolic geometry of complex networks},
  author    = {Krioukov, Dmitri and Papadopoulos, Fragkiskos and Kitsak, Maksim and Vahdat, Amin and Bogu{\~n}{\'a}, Mari{\'a}n},
  journal   = {Physical Review E},
  volume    = {82},
  number    = {3},
  pages     = {036106},
  year      = {2010},
  doi       = {10.1103/PhysRevE.82.036106},
  url       = {https://doi.org/10.1103/PhysRevE.82.036106},
  publisher = {American Physical Society}
}

@article{Boguna2,
  title     = {Sustaining the internet with hyperbolic mapping},
  author    = {Bogu{\~n}{\'a}, Mari{\'a}n and Papadopoulos, Fragkiskos and Krioukov, Dmitri},
  journal   = {Nature Communications},
  volume    = {1},
  number    = {1},
  pages     = {62},
  year      = {2010},
  doi       = {10.1038/ncomms1063},
  url       = {https://doi.org/10.1038/ncomms1063},
  publisher = {Nature Publishing Group}
}

@incollection{akkermans,
  author       = {Omrie Ovdat and Eric Akkermans},
  title        = {The breaking of continuous scale invariance to discrete scale invariance: a universal quantum phase transition},
  booktitle    = {Fractal Geometry and Stochastics VI},
  editor       = {Uta Freiberg and Ben Hambly and Michael Hinz and Steffen Winter},
  series       = {Progress in Probability},
  volume       = {76},
  pages        = {209--238},
  publisher    = {Birkhäuser Cham},
  year         = {2021},
  doi          = {10.1007/978-3-030-59649-1_9}
}

@article{hyp-efimov1,
  author  = {Nabil Iqbal and Hong Liu and M\'ark Mezei and Qimiao Si},
  title   = {Quantum phase transitions in holographic models of magnetism and superconductors},
  journal = {Phys.\ Rev.\ D},
  volume  = {82},
  number  = {045002},
  pages   = {045002},
  year    = {2010},
  doi     = {10.1103/PhysRevD.82.045002}
}

@article{hyp-efimov2,
  author       = {Ren Zhang and Chenwei Lv and Yangqian Yan and Qi Zhou},
  title        = {Efimov-like states and quantum funneling effects on synthetic hyperbolic surfaces},
  journal      = {Science Bulletin},
  volume       = {66},
  number       = {19},
  pages        = {1967--1972},
  year         = {2021},
  doi          = {10.1016/j.scib.2021.06.017}
}

@book{newmanbook,
    author = {Newman, Mark},
    title = {Networks},
    publisher = {Oxford University Press},
    year = {2018},
    month = {07},
    isbn = {9780198805090},
    doi = {10.1093/oso/9780198805090.001.0001},
    url = {https://doi.org/10.1093/oso/9780198805090.001.0001},
}

@article{new-localization,
    author = {Avetisov, V and Gorsky, A and Nechaev, S and Valba, O},
    title = {Localization and non-ergodicity in clustered random networks},
    journal = {Journal of Complex Networks},
    volume = {8},
    number = {2},
    pages = {cnz026},
    year = {2020},
    month = {04},
    issn = {2051-1329},
    doi = {10.1093/comnet/cnz026},
    url = {https://doi.org/10.1093/comnet/cnz026}
}

@article{Kenyon_2017-1,
  author    = {Richard Kenyon and Charles Radin and Kui Ren and Lorenzo Sadun},
  title     = {Multipodal Structure and Phase Transitions in Large Constrained Graphs},
  journal   = {Journal of Statistical Physics},
  year      = {2017},
  volume    = {168},
  number    = {2},
  pages     = {233--258},
  doi       = {10.1007/s10955-017-1804-0},
  url       = {https://doi.org/10.1007/s10955-017-1804-0}
}

@article{Kenyon_2017-2,
doi = {10.1088/1751-8121/aa8ce1},
url = {https://doi.org/10.1088/1751-8121/aa8ce1},
year = {2017},
month = {oct},
publisher = {IOP Publishing},
volume = {50},
number = {43},
pages = {435001},
author = {Kenyon, Richard and Radin, Charles and Ren, Kui and Sadun, Lorenzo},
title = {The phases of large networks with edge and triangle constraints},
journal = {Journal of Physics A: Mathematical and Theoretical}
}

@article{Gorsky:2020cnaa008,
  author  = {Gorsky, A. and Valba, O.},
  title   = {Finite-size effects in exponential random graphs},
  journal = {Journal of Complex Networks},
  volume  = {8},
  number  = {1},
  pages   = {cnaa008},
  year    = {2020},
  doi     = {10.1093/comnet/cnaa008},
  eprint  = {1905.03336},
  archivePrefix = {arXiv},
  primaryClass  = {cond-mat.dis-nn}
}

@article{Gorsky:2021prd,
  author  = {Gorsky, A. and Valba, O.},
  title   = {Interacting thermo-field doubles and critical behavior in random regular graphs},
  journal = {Physical Review D},
  volume  = {103},
  pages   = {106013},
  year    = {2021},
  doi     = {10.1103/PhysRevD.103.106013},
  eprint  = {2101.04072},
  archivePrefix = {arXiv},
  primaryClass  = {hep-th}
}

@article{Kochergin:2023prb,
  author  = {Kochergin, Daniil and Khaymovich, Ivan M. and Valba, Olga and Gorsky, Alexander},
  title   = {Anatomy of the fragmented Hilbert space: eigenvalue tunneling, quantum scars and localization in the perturbed random regular graph},
  journal = {Physical Review B},
  volume  = {108},
  pages   = {094203},
  year    = {2023},
  doi     = {10.1103/PhysRevB.108.094203},
  eprint  = {2305.14416},
  archivePrefix = {arXiv},
  primaryClass  = {cond-mat.dis-nn}
}

@article{PArev,
author = {Michael Mitzenmacher},
title = {{A Brief History of Generative Models for Power Law and Lognormal Distributions}},
volume = {1},
journal = {Internet Mathematics},
number = {2},
publisher = {A K Peters, Ltd.},
pages = {226 -- 251},
year = {2003},
}

@article{Trugenberger_2017, 
title={Combinatorial quantum gravity: geometry from random bits}, 
volume={2017}, 
DOI={10.1007/jhep09(2017)045}, 
number={9}, 
journal={Journal of High Energy Physics}, 
publisher={Springer Science and Business Media LLC}, 
author={Trugenberger, Carlo A.}, 
year={2017}
}

@article{Trugenberger_2025, 
title={Networks as the fundamental constituents of the Universe}, 
volume={6}, 
DOI={10.1088/2632-072x/ae29d3}, 
number={4}, 
journal={Journal of Physics: Complexity}, 
publisher={IOP Publishing}, 
author={Trugenberger, C A}, 
year={2025}, 
pages={042001}
}

@article{Broido_Clauset, 
title={Scale-free networks are rare}, 
volume={10}, 
DOI={10.1038/s41467-019-08746-5}, 
number={1}, 
journal={Nature Communications}, 
publisher={Springer Science and Business Media LLC}, 
author={Broido, Anna D. and Clauset, Aaron}, 
year={2019}
}

@article{Caldarelli,
author = {Matteo Serafino  and Giulio Cimini  and Amos Maritan  and Andrea Rinaldo  and Samir Suweis  and Jayanth R. Banavar  and Guido Caldarelli },
title = {True scale-free networks hidden by finite size effects},
journal = {Proceedings of the National Academy of Sciences},
volume = {118},
number = {2},
pages = {e2013825118},
year = {2021},
doi = {10.1073/pnas.2013825118},
URL = {https://www.pnas.org/doi/abs/10.1073/pnas.2013825118},
eprint = {https://www.pnas.org/doi/pdf/10.1073/pnas.2013825118},
}

@article{maslov,
  author    = {Sergei Maslov and Kim Sneppen},
  title     = {Specificity and Stability in Topology of Protein Networks},
  journal   = {Science},
  volume    = {296},
  number    = {5569},
  pages     = {910--913},
  year      = {2002},
  doi       = {10.1126/science.1065103},
  url       = {https://doi.org/10.1126/science.1065103}
}

@inproceedings{viger,
  author    = {Fabien Viger and Matthieu Latapy},
  title     = {Efficient and simple generation of random simple connected graphs with prescribed degree sequence},
  booktitle = {Computing and Combinatorics (COCOON 2005)},
  series    = {Lecture Notes in Computer Science},
  volume    = {3595},
  pages     = {440--449},
  year      = {2005},
  publisher = {Springer},
  address   = {Springer, Berlin, Germany},
  doi       = {10.1007/11510290_44}
}

@article{sia2019ollivier,
  title={Ollivier-Ricci Curvature-Based Method to Community Detection in Complex Networks},
  author={Sia, Jayson and Jonckheere, Edmond and Bogdan, Paul},
  journal={Scientific Reports},
  volume={9},
  number={1},
  pages={9800},
  year={2019},
  publisher={Nature Publishing Group},
  doi={10.1038/s41598-019-46079-x},
  url={https://doi.org/10.1038/s41598-019-46079-x}
}

@book{kinetics,
  author    = {Laidler, Keith J.},
  title     = {Chemical Kinetics},
  edition   = {3rd},
  publisher = {Harper \& Row},
  year      = {1987},
  address   = {New York}
}

@article{Srinivasan2025,
  author  = {Srinivasan, Bhargavi},
  title   = {Intrinsic Geometry of the Stock Market from Graph {R}icci Flow},
  journal = {arXiv preprint arXiv:2510.15942},
  year    = {2025}
}

@article{BAmodel,
  title = {Emergence of {{Scaling}} in {{Random Networks}}},
  author = {Barab{\'a}si, Albert-L{\'a}szl{\'o} and Albert, R{\'e}ka},
  year = 1999,
  month = oct,
  journal = {Science},
  volume = {286},
  number = {5439},
  pages = {509--512},
  publisher = {American Association for the Advancement of Science},
  doi = {10.1126/science.286.5439.509},
  urldate = {2026-11-25}
}

@article{JostLiu2014,
  author  = {Jost, J{\"u}rgen and Liu, Shiping},
  title   = {Ollivier's {R}icci curvature, local clustering and curvature-dimension inequalities on graphs},
  journal = {Discrete \& Computational Geometry},
  volume  = {51},
  number  = {2},
  pages   = {300--322},
  year    = {2014}
}

@article{Bhattacharya2015,
  author  = {Bhattacharya, Bhaswar B. and Mukherjee, Sumit},
  title   = {Exact and asymptotic results on coarse {R}icci curvature of graphs},
  journal = {Discrete Mathematics},
  volume  = {338},
  number  = {1},
  pages   = {23--42},
  year    = {2015}
  }

\end{document}